\documentstyle[preprint,aps]{revtex}

\begin{document}

\draft

\title{Numerical Studies of the two-leg Hubbard ladder
}
\author{Zheng Weihong$^1$\cite{ezwh}, J. Oitmaa$^1$\cite{eotja}, C.J. 
Hamer$^1$\cite{ecjh},
and R.J. Bursill$^2$\cite{erjb}
}
\address{
$^1$School of Physics, The University of New South Wales, Sydney, 
NSW 2052, Australia.\\
$^2$Department of Physics, UMIST, PO Box 88, Manchester, M60 1QD, UK.
}

\date{\today}

\maketitle

\begin{abstract}
The Hubbard model on a two-leg ladder structure has been studied by a 
combination of series expansions at $T=0$ and the density-matrix 
renormalization group. We report results for the ground state energy 
$E_0$ and spin-gap $\Delta_s$ at half-filling, as well as dispersion 
curves for one and two-hole excitations. For small $U$ both $E_0$ and 
$\Delta_s$ show  a dramatic drop near $t/t_{\perp}\sim 0.5$, which 
becomes more gradual for larger $U$. This represents a crossover from a 
``band insulator" phase to a strongly correlated spin liquid. 
The lowest-lying two-hole state rapidly becomes strongly bound as $t/t_{\perp}$
increases, indicating the possibility that phase separation may occur.
The various features are collected in a 
``phase diagram" for the model.
\end{abstract}



\narrowtext

\section{Introduction}

\label{intro}

The last decade has seen a great deal of interest in spin and/or 
correlated electron systems on a ladder structure formed from two 
coupled chains. This work has been motivated both by the discovery of 
real materials with $S=\frac{1}{2}$ ions forming a ladder 
structure\cite{azu94}, and because ladder systems  exhibit a number of 
interesting and surprising properties (see e.g.\ the review of Dagotto 
and Rice\cite{dag96}).

Particularly interesting behaviour may be expected if the system 
includes charge degrees of freedom. This can be achieved by doping, to 
create a system of strongly correlated mobile holes, as in the cuprate 
superconductors. 
Systems studied include LaCuO$_{2.5}$ doped with Sr\cite{azu94}
and Ca$_{14}$Cu$_{24}$O$_{41-\delta}$ doped with Sr\cite{ueh96},
the latter showing superconductivity under pressure.
The $t$-$J$ and Hubbard models provide alternate 
representations of the physics of such systems, and both have been 
studied extensively. References to most of the existing work on the 
$t$-$J$ ladder system, as well as the most recent evidence for the form 
of the phase diagram can be found in M\"uller and Rice\cite{mul98}.

We are here interested in the repulsive ($U>0$) Hubbard model on a 2-leg 
ladder. The first study of this system, to our knowledge, was by 
Fabrizio, Parola and Tosatti\cite{fab92} who used a weak-coupling 
renormalization group approach to investigate the role of the interchain 
hopping $t_{\perp}$ in driving the system out of a Luttinger liquid 
phase. Earlier work, in this context, had considered a 2-dimensional 
system of weakly coupled chains\cite{bou91}. Further 
work\cite{khv94,bal96}, using bosonization techniques, valid for weak 
interchain coupling, has identified a number of possible phases which 
the 2-leg Hubbard ladder may exhibit. In the notation of Balents and 
Fisher\cite{bal96}, these are denoted as C$n$S$m$ where $n$, $m$ 
represent the number of gapless charge and spin modes respectively. (Here 
$n,m=0,1,2$ giving 9 possible phases).

Numerical studies of the Hubbard ladder have been carried out for both 
static and dynamic properties. The Density Matrix Renormalization Group 
(DMRG) technique\cite{whi93} has been used to calculate both spin and 
pairing correlations\cite{noa94,noa96}, and to calculate the spin and 
charge gaps\cite{noa96,par99} as functions of the parameters of the 
model. The DMRG technique, at least in its present form, is unable to 
compute the dynamic response, particulary momentum dependent properties. 
The one-hole spectral function has been obtained using exact 
diagonalizations\cite{rie99} (limited to $2\times 8$ sites) and quantum 
Monte Carlo techniques\cite{end96}. This latter paper also reports 
results for two-hole spin and charge excitations.
The spin-gap at half-filling, obtained by DMRG on a $2\times 32$
lattice\cite{noa96} showed a very sudden decrease near $t/t_{\perp}\simeq
0.5$, particularly for small $U$. However finite size effects are present
and may mask the true behaviour. This work was motivated in part by a need to 
confirm this, and in fact to look for possible evidence of a phase transition
near this point, as well as by a desire to explore the form of
dispersion curves for spin and charge excitations, which cannot be obtained by
DMRG methods.

The Hamiltonian of the Hubbard ladder  is written as
\begin{eqnarray}
H = - t \sum_{i,a,\sigma}  (c^{\dag}_{i,a,\sigma} c_{i+1,a,\sigma}         
+ H.c. )   - t_{\perp} \sum_{i,\sigma} (c^{\dag}_{i,1,\sigma} 
c_{i,2,\sigma}         + H.c. )   + U \sum_{ia} n_{i,a,\uparrow} 
n_{i,a,\downarrow}         
\label{eqH}
\end{eqnarray}
where $i$ labels the rungs of the ladder, $a$ (=1,2) is a leg index, 
$\sigma$ ($=\uparrow$, $\downarrow$) is the spin index, and the 
operators have the usual meaning. 
There are several instructive limiting cases. When $U=0$ the electrons 
are non-interacting and the system has two simple cosine bands of width 
$4t$ and separation $2 t_{\perp}$. The system is metallic for all 
electron densities $n$, except for the case $n=1$ (half-filling) and 
$t_{\perp}>2t$ when there is a band gap and the lower (bonding) band is 
completely filled. The case is aptly referred to as a ``band 
insulator"\cite{noa96}. On the other hand when the interchain hopping 
$t_{\perp}=0$ (and $U>0$) the ladder decouples to two Hubbard chains for 
which there are exact, but highly nontrivial solutions. The system is 
then in a Luttinger liquid phase, and also an insulator at half-filling.

When both $t_{\perp}$ and $U$ are non-zero there are no exact results 
known. However the form of the phase diagram can be reasonably inferred 
from the analytic and numerical calculations referred to above. A nice 
discussion is given by Noack {\it et al.}\cite{noa96}. In the limit of 
large $U$ doubly occupied sites are suppressed and the model reduces to 
a $t$-$J$ ladder, with parameters $J= 2 t^2/U$, $J_{\perp} = 2 
t_{\perp}^2/U$. The Hubbard ladder has electron hole symmetry under the 
transformation
\begin{equation}
c_{i,a,\sigma} \to 1 - c_{i,a,\sigma}
\end{equation}
The phase diagram and properties of the model are thus symmetric about 
the $n=1$ case. We see manifestations of this in the series for various 
excitation energies, discussed below.

In this paper we study the Hubbard ladder using both the method of 
series expansions\cite{gel90,he90,gelfand} and DMRG\cite{whi93}. The 
series expansion method is complementary to other numerical methods and 
is able to provide ground state properties, excitation spectra and $T=0$ 
critical points to high accuracy. Another advantage is that one deals 
with a system in the thermodynamic limit and finite size corrections are 
not needed. We have used this approach recently in studies of the 
$t$-$J$ model on the square lattice\cite{ham98} and on a 2-leg 
ladder\cite{oit99}, and we refer to those papers for details of the 
method and references to previous work. The emphasis of our work is on 
the half-filled case $n=1$, and on one and two-hole excitations from 
half-filling. The series method is not well suited to handle variable 
electron density.

We are not aware of any previous series work on the Hubbard ladder. Shi 
and Singh\cite{shi95} have used a somewhat different series approach to 
study the Hubbard model on the square lattice. Their method,  which 
introduces an artificial antiferromagnetic Ising term into the 
Hamiltonian is not appropriate here, as we do not expect any magnetic 
long range order in the ladder system.

This paper is organized as follows. In Sec. II, we discuss briefly the 
methods used. In Sec. III, we study the system at half-filling. In Sec. 
IV, we consider the system with one and two holes. The last section is 
devoted to discussion and conclusions.

\section{Methods}


The series expansion method is based on a linked cluster formulation of 
standard Rayleigh-Schr\"odinger perturbation theory. We use a ``rung 
basis" and write the Hamiltonian in the form
\begin{equation}
H = H_0 + x V
\end{equation}
where
\begin{eqnarray}
H_0 &=& - t_{\perp} \sum_{i,\sigma}  (c^{\dag}_{i,1,\sigma} 
c_{i,2,\sigma} + H.c. )  + U \sum_{ia} n_{i,a,\uparrow} 
n_{i,a,\downarrow}
\end{eqnarray}
is taken to be the unperturbed Hamiltonian and
\begin{eqnarray}
x V &=& - t \sum_{i,a,\sigma}  (c^{\dag}_{i,a,\sigma} c_{i+1,a,\sigma}         
+ H.c. )
\end{eqnarray}
is the perturbation. Thus $H_0$ describes decoupled rungs, and can be 
solved exactly while the intrachain hopping term couples the rungs and 
is treated perturbatively.

The Hamiltonian for a single rung has 16 possible states. These are 
shown in Table I. 
For $U<3t_{\perp}$ the  lowest energy 
rung state is a spin-singlet containing two electrons (state 0 in Table 
I). At $U=3 t_{\perp}$ there is a level crossing and the lowest energy 
rung state becomes a doublet $S=\frac{1}{2}$ state with a single electron 
in a symmetric (bonding) state for $U>3 t_{\perp}$. The eigenstates of 
$H_0$ are then direct products constructed from the possible rung 
states. The ground state of $H_0$ for the half-filling system ($n=1$) is 
the state that has each rung in a spin singlet. This is true even for 
$U>3 t_{\perp}$, as transferring an electron from a doubly occupied rung 
to another doubly occupied rung costs energy.

To compute the perturbation series we fix the values of $t_{\perp}$ and 
$U$ and expand in powers of $x\equiv t/t_{\perp}$. Without loss of 
generality we set $t_{\perp}=1$ to define the energy scale. The series 
are then evaluated at the desired value of $t$ using standard Pad\'e 
approximants and integrated differential approximants\cite{gut}.

We have also carried out large-scale DMRG calculations\cite{whi93}. Two 
different DMRG algorithms have been employed. Both are `infinite lattice' 
algorithms \cite{whi93} with open boundary conditions. The first method 
uses a superblock consisting of the usual system and environment blocks 
with two added rungs in the middle. The system/environment blocks are 
augmented by one rung at a time and the superblocks always have an even 
number of rungs. The second method is similar except that only one rung 
is kept in the middle, meaning that the superblocks have an odd number 
of rungs. The second method allows more states to be retained in the 
blocks, whilst the even lattices dealt with in the first method are 
usually considered to be more desirable for finite-size scaling (FSS) studies. 
For the first method typical calculations involve ladders with up to 60 
rungs, keeping up to 550 states per block. For the second method the 
superblocks studied typically reached 81 rungs and up to 1500 states
were retained per block.
It should be noted that the `infinite lattice' algorithm, despite
its name, can be used to obtain accurate results for finite lattices,
as we shall show in what follows.

\section{The Half-Filled Case}

\subsection{Ground state energy}

We first consider the ground state energy $E_0$ at half-filling. The 
series method yields an expansion of
\begin{equation}
E_0 = \sum_n a_n (U/t_{\perp}) (t/t_{\perp})^n
\end{equation}
The series have been computed to order $x^{14}$ for various 
$U/t_{\perp}$. The coefficients for $U / t_{\perp} = 8$ are given in 
Table II, and the other coefficients can be provided on request. The 
cluster data for this one-dimensional problem are trivial. The limiting 
factor is the size of the matrices used in obtaining the cluster 
energies. DMRG gives the ground state energy directly for a given 
lattice, and finite-size scaling must be used to extract the bulk 
result.

In Figure \ref{fig_2} we show our results for the ground state energy in 
the half-filled case versus $t$ for various $U/t_{\perp}$ obtained both 
by series expansions and DMRG calculations.
For given $t$ the energy increases with $U$ as expected. For fixed $U$ 
the energy decreases slowly with increasing $t$. The series are well 
converged for $t/t_{\perp} \lesssim 0.6$, but the convergence 
deteriorates rapidly at that value. The DMRG is, however, well converged 
even for larger $t$ and agrees well with the series results for smaller 
$t$. A remarkable feature of the ground state energy is the sharp downturn 
which occurs near $t/t_{\perp}\sim 0.5 - 0.6$ for $U/t_{\perp} 
\lesssim 1$, but which becomes smoother for larger $U$. For the case of 
free electrons ($U=0$) the ground state energy is exactly
\begin{equation}
{E_0\over Nt_{\perp}} = \cases {-1, & $t/t_{\perp}< 1/2$; \cr  -1 - 
{2\over \pi} {\Big [}  \sqrt{ (2t/t_{\perp})^2-1} - \cos^{-1} {\Big (} 
{t_{\perp} \over 2 t} {\Big )} {\Big ]}, & $t/t_{\perp}> 1/2$.\cr }
\end{equation}
and, as mentioned above, the crossover represents a transition from band 
insulator to a conductor with gapless charge and spin excitations. The 
qualitative similarity between the $U=0$ case and small $U$ (e.g.\ 
$U/t_{\perp}=0.5$ in the figure) suggests that spin excitations may 
remain gapless, or nearly so, for small finite $U$ and large $t$.

We can compare our results for the ground state energy with those of Kim 
{\it et al.}\cite{kim99}, obtained via a variational method. These are shown
in Figure 2. The results are in good agreement for small $t/t_{\perp}$,
but it appears that the variational method significantly over estimates
the ground state energy for large $t/t_{\perp}$.



\subsection{Spin excitations}

We now turn to the excitations, which  we have computed 
directly using Gelfand's method\cite{gelfand}. The lowest energy spin 
excitation branch results from exciting a rung from a spin singlet 
(state 0 in Table I) to a triplet (any of states 5,6,7 in Table I). 
Nonzero $t$ allows this to propagate coherently along the chain, giving 
rise to a triplet magnon excitation. Figure \ref{fig_3} shows dispersion 
curves $\Delta_s (k)$ for this excitation for fixed $t/t_{\perp}=0.5$ 
and various $U/t_{\perp}$. For larger $t/t_{\perp}$ the series become 
too erratic. For all cases the energy minimum occurs at $k=\pi$. As $U$ 
increases the  
 bandwidth of the spin excitation 
 decreases.

The excitation energy at $k=\pi$ defines the spin gap 
$\Delta_s \equiv \Delta_s(\pi)$. 
The raw data are shown in Figure \ref{fig_4}, where we plot 
$\Delta_s / t_\perp$ versus $t/t_{\perp}$ for various $U/t_{\perp}$. For 
small $t$ our series give rather precise estimates. In the limit $t=0$ 
the spin gap is $\frac{1}{2} (\sqrt{U^2 + 16 t_{\perp}^2 } - U )$ which is 
$2t_{\perp}$ for $U=0$ and decreases for increasing $U$. Near 
$t/t_{\perp} \sim 0.4$--0.5 there is a crossover and beyond this point 
the spin gap for large $U$ exceeds that for small $U$. 
For $U/t_{\perp}\lesssim 2$ this crossover becomes very sharp.
The series extrapolations would indicate that the spin gap actually vanishes at
$t/t_{\perp}\simeq 0.5-0.6$, but we believe this is an artefact
of poor convergence in this region. It is expected\cite{bal96,noa96} 
 that at half-filling
the system will have a finite spin gap throughout the phase diagram
for any $U>0$, due to `umklapp' processes.
It is clear however that the spin gap 
becomes quite small for $t/t_{\perp} \gtrsim 0.5$.


We have also used our DMRG algorithms to calculate $\Delta_s$. In 
order to obtain insight into the potential limitations of DMRG methods 
for this problem, we consider the finite-size scaling of 
$\Delta_s$ in the exactly solvable $U = 0$ 
case. As mentioned, for $U = 0$, the system is gapped for 
$t_{\perp} > 2 t$ and gapless for $t_{\perp} < 2 t$. In Fig.\ 
\ref{fig_5a} we plot $\Delta_s$ as a function of $1 / N_{\rm rung}$ for 
three values of the ratio $t / t_\perp$. For $t / t_\perp = 0.49$ (just 
below the critical ratio), the system has a small spin gap, and 
$\Delta_s$ scales smoothly to its bulk value, the finite-size 
corrections vanishing as $1 / N_{\rm rung}^2$ (as is to be expected for 
a gapped system with open boundary conditions). For 
$t / t_\perp = 0.51$ (just above the critical ratio), the gap suddenly 
begins to display a ``sawtooth" dependence on $N_{\rm rung}$, with
the points of the saw scaling towards zero as $N_{\rm rung}\to \infty$.
This behaviour becomes more pronounced as we move further beyond the
critical ratio. It appears that multiple crossovers are occurring,
with new states crossing over to the bottom of the spectrum as
$N_{\rm rung}$ increases. These sawtooth oscillations are presumably
due to the 2-band structure: every so often it is is energetically
favorable for an extra electron to be added to the top band rather than the
bottom band, and a crossover occurs. This
oscillatory behaviour hs not been exhibited before, as far as we are aware.

In order to calculate the bulk spin gap successfully with DMRG, we must 
first accurately calculate $\Delta_s$ for a number of lattice sizes 
$N_{\rm rung}$, and then extrapolate to the bulk limit, assuming some 
scaling ansatz for the finite $N_{\rm rung}$ corrections. 
An example of this is shown in 
Table \ref{table_3}, where we compare DMRG and exact results for 
$\Delta_s$ for finite lattices and the bulk limit. In this case, where 
the FSS is smooth, it is possible to obtain a reasonable estimate for 
the bulk spin gap, the real error being around 0.1\%. 
Two potential 
problems emerge from our studies of gapless systems in the $U = 0$ case
for $t/t_{\perp}>0.5$. 
Firstly, because of the oscillating FSS 
it might be difficult to obtain 
accurate DMRG results for very large lattices in a regime where the
gap is small or vanishing. Convergence of finite lattice results 
with the number of states, $m$, retained per block, must be monitored 
over a large range of $m$ values. For a given lattice size, improved 
DMRG estimates can be obtained by using a finite lattice algorithm 
\cite{whi93}. However, even if highly accurate results are available for 
a number of lattice sizes, a second and more pressing problem is that, 
as a result of the oscillations or erratic FSS, extremely large lattices 
may be needed in order to reach a regime where a suitable scaling ansatz 
can be reliably used to extrapolate to the bulk limit, as can be seen 
for the gapless cases in Fig.\ \ref{fig_5a}.

Fortunately, the presence of electron repulsion smooths out the FSS, 
allowing reliable DMRG estimates of $\Delta_s$ some way beyond 
$t / t_\perp = 0.5$. In Fig.\ \ref{fig_5}, we show the FSS of DMRG 
estimates of $\Delta_s$ in the $U / t_\perp = 1$ case for various 
$t / t_\perp$. We find that up to around $t / t_\perp = 0.55$, the 
corrections to the bulk results for $\Delta_s$ scale as $1 / N_{\rm 
rung}^2$, as in the gapped case for $U = 0$. As $t / t_\perp$ is 
increased, however, we observe initially oscillatory or erratic FSS, 
followed by a crossover to linear dependence of $\Delta_s$ on $1 / 
N_{\rm rung}$, as might be expected for a system with a small gap. This 
is illustrated in Fig.\ \ref{fig_5} for the $t / t_\perp = 0.6$ case, 
where a crossover to linear behaviour is observed as the lattice size 
reaches around 30 rungs. As mentioned, it is important to first assess 
the DMRG convergence for the finite lattices before attempting 
extrapolations. In Table \ref{table_4}, we show the DMRG convergence of 
$\Delta_s$ with $m$ for the odd-rung algorithm, with $m$ ranging from 
200 to 1500. It can be seen in this case that the finite lattice 
estimates are sufficiently well converged to afford reliable 
extrapolations. In Fig.\ \ref{fig_5} results from the odd-rung algorithm 
with $m = 750$ are plotted along with results from the even-rung 
algorithm with $m = 800$. Good agreement can be seen between the two 
sets of results in the linear regime. In order to make bulk estimates 
for $t / t_\perp \geq 0.6$, we assume a linear scaling ansatz. 
Presumably, if the resulting estimate is non-zero, as is the case, e.g., 
for $t / t_\perp = 0.6$, there should be a crossover from linear to 
quadratic scaling for larger lattices still, in which case the estimates 
are (tight) lower bounds on the spin gap. For
the case of $U / t_\perp = 1$, we 
can carry this out up until $t / t_\perp$ reaches around 0.75. Beyond 
this value, the FSS is too erratic to permit accurate finite lattice 
estimates of the spin gap for sufficiently large lattices. In the inset 
to Fig.\ 5 this erratic FSS is shown for the $t / t_\perp = 0.8$ case.

Plots of the DMRG bulk estimates of $\Delta_s$ as functions of 
$t / t_\perp$ are included in Fig.\ \ref{fig_4}. The spin gap obtained 
in this way is indistinguishable from the series values for small $t$. 
However, the DMRG also provides well converged results for larger $t$ 
provided that $U$ is not too small.
In the $U / t_\perp = 2$ case, for example, $\Delta_s$ undergoes a very 
rapid decrease up till $t/t_{\perp}\simeq 0.6$, and
then flattens out at a small but finite value.
The same behaviour was already seen in the previous DMRG calculation of
Noack {\it et al.}\cite{noa96}.
Noack {\it et al.}\cite{noa96} did their calculation at a fixed lattice size
$2\times 32$ sites, however, whereas we performed a careful finite-size
scaling analysis to confirm that the spin gap remains finite, rather than
scaling to zero in the bulk limit.

In the case $U/t_{\perp}=1$ case, a similar
effect occurs, but the FSS behaviour becomes completely erratic and 
the extrapolations fail at 
around $t / t_\perp \approx 0.75$. In Fig.\ \ref{fig_6} we depict the 
region in the $(U / t_\perp)$-$(t / t_\perp)$ plane where the DMRG either 
fails to give reasonable results due to erratic FSS, or indicates a 
very small or vanishing gap, by marking it with an ``F''. Also included 
in Fig.\ \ref{fig_6} is the line where the series appear to give a vanishing 
spin gap. This line indicates the crossover in the physics of the system from 
band-insulator to strongly correlated Mott insulator.


\section{1 and 2-hole states}

As mentioned in Section \ref{intro} there is considerable interest 
in the doped Hubbard ladder where the electron density $n<1$. The series 
method is not well suited to study the effect of finite doping. However 
we are able to compute the ground state and excitation energies when the 
system contains one or two holes.

\subsection{1-Hole Case}

In the $t=0$ limit a single hole will change one of the singlet rung 
states into an $S^z =\frac{1}{2}$ or $-\frac{1}{2}$ bonding state, with an 
energy increase of $-t_{\perp}-\lambda_1$. Finite $t$ will allow the 
hole to propagate along the ladder, giving a quasiparticle band. Figure 
\ref{fig_7} shows the quasiparticle excitation energy $\Delta_{\rm 
1h}/t_{\perp}$ as a function of wavenumber for the case 
$t/t_{\perp}=0.5$ and various $U/t_{\perp}$. For $U=0$ we have the exact 
results
\begin{equation}
\Delta_{\rm 1h}(k) = t_{\perp} +  2 t \cos(k)
\end{equation}
and this is seen through the vanishing of all higher terms in the 
series. For the choice $t/t_{\perp}=0.5$ this gives $\Delta_{\rm 1h}=0$ 
at $k=\pi$. We also show, for comparison, the approximate dispersion 
curve for $U/t_{\perp}=4$ obtained by Endres {\it et al.}\cite{end96}, 
through their ``local rung approximation". This appears to overestimate 
the energy by about $0.4 t_{\perp}$, although the overall shape is very 
similar.


For increasing $U$ the energies are depressed and there is a small 
decrease in the quasiparticle bandwidth although the overall cosine 
shape remains. The minimum of the quasiparticle spectrum occurs at 
$k=\pi$ throughout. The energy zero is taken to be the ground state 
energy of the half-filled ladder. Each quasiparticle band crosses the 
zero level, indicating that the overall energy of the system is reduced 
when a hole is created, and this is also marked in Fig.\ \ref{fig_6}. 
Figure \ref{fig_8} shows the ``quasiparticle gap" as a function of 
$t/t_{\perp}$ for various $U/t_{\perp}$ obtained from the series and 
from DMRG. Both methods agree up to $t/t_{\perp}\sim 0.5$, beyond which 
the series fail to converge. 
This is the same crossover region seen in the spin gap studies.
The DMRG results suggest an upturn or change in slope 
 for larger $t/t_{\perp}$ and not very large $U$, as shown in the 
Figure for $U=1$. However this remains somewhat speculative as the 
scaling behaviour of the DMRG estimates is unusual. We illustrate this 
in Figure \ref{fig_9}, where we plot the gap (for $U/t_{\perp}=1$) 
versus $1/N^2$. The behaviour shows a qualitative change between 
$t/t_{\perp}=0.5$ and 0.6 becoming oscillatory and with a change of sign 
of limiting slope. Nevertheless the larger $N$ points scale quite well.
Noack {\it et al.}\cite{noa96} saw s similar break in the behaviour of the 
charge gap at the crossover for $U/t_{\perp}=4$.


\subsection{2-Hole Case}

We would also like to explore the system doped with two holes,
to see if binding occurs between the holes. From Table I,
one can see that at zeroth order ($t=0$), the energy gap for two
holes sitting on the same rung is ${\case 1/2}(\sqrt{U^2 + 16 t_{\perp}^2 } - U)$,
which is larger than the gap
$(\sqrt{U^2 + 16 t_{\perp}^2 } - U)-2t_{\perp}$ for two holes on different
rungs. Thus it is not energetically favourable to have two holes on the same rung.
Our present series methods, unfortunately, are unable to treat the latter case 
of two holes on separate rungs: so here we restrict ourselves to exploring
whether the former state becomes bound at larger $t$.
One starts from a state with both holes on a single rung and all other 
rungs in spin singlet states; and then the hopping term allows these holes 
to move along the ladder, and so generates a dispersion relation for this state.
For example at $U/t_{\perp}=8$,
our second order series result for the dispersion is
\begin{equation}
\Delta_{2h} (k)/t_{\perp} = 0.4721 + ( 1.805 + 1.447 \cos k ) (t/t_{\perp})^2
\end{equation}
One can see that the  minimum energy is at
$k=\pi$, rather than $0$,  due to  the fact that this is not 
the lowest energy two-hole state.
Comparing with the one-hole dispersion relation, we find
no evidence for binding of this state at higher $t$.

Our series results do not preclude the existence of a two-hole bound state of more 
complex structure, and there are indications from other work that this 
can occur. Noack {\it et al.}\cite{noa94,noa96} find, for a $2\times 32$ 
lattice, a small binding energy ($E_b \simeq 0.14$) for two holes, and a 
pair wavefunction which has roughly equal amplitudes on a rung and 
between nearest neighours along the same leg of the ladder. Finite size 
effects may be large enough to mask the true behaviour. Kim {\it et 
al.}\cite{kim99} also discuss pair formation using both DMRG and a 
variational method. They conclude that, at least for large $U$, holes 
favour adjacent rungs to minimise the Coulomb energy. Our series method 
is unable to explore such complex pair states. 

Instead, we have used the DMRG 
method to compute the minimum energy of the system with one and two 
holes, up to ladders of size $2\times L$ ($L=60$), and hence the binding energy 
defined by
\begin{equation}
E_b = 2 [E_0 (L,L-1) - E_0(L,L)] - E_0 (L-1,L-1) + E_0(L,L) 
%
\end{equation}
where $E_0(N_{\uparrow}, N_{\downarrow})$ is the ground state energy 
with $N_{\uparrow}$ ($N_{\downarrow}$) up (down) electrons.
 We show 
this as a function of $t/t_{\perp}$ for $U/t_{\perp}=2,8$ in Figure 
\ref{fig_12}. As can be seen, there is no binding for small $t/t_{\perp}$, 
and the binding energy is zero, corresponding to an unbound,
well seperated pair of holes.
%
For $t/t_{\perp}$ larger than a critical value $(t/t_{\perp})_c$, however, 
 hole binding becomes energetically favourable.
In fact as we can see from  Figure 
\ref{fig_12}, $E_b$ increases from zero
extremely quickly once past the critical $(t/t_{\perp})_c$.
Now the binding of two holes is expected to be a necessary though not
sufficient precondition for the phenomenon of phase separation (binding of many
holes). Figure \ref{fig_12} leads us suppose that phase separation may well
occur for large $t/t_{\perp}$; or else at
fixed $t/t_{\perp}=1$, phase separation may well occur at small $U/t_{\perp}$.
This would accord with our current knowledge of the $t-J$ ladder, where
the exchange coupling
$J\sim 4t^2/U$, and phase separation occurs\cite{rom00}
at small $U/t$, i.e. small $t/J$.
The boundary for two-hole binding is shown in our ``phase diagram", Figure \ref{fig_6}.
No explicit search for phase separation has yet made in the Hubbard ladder,
as far as we are aware.

\section{Discussion and Conclusions}

We have made the first application of the series expansion method to the
Hubbard model on a two-leg ladder, and also used the DMRG method to explore 
its properties  at 
half-filling at $T=0$. Our series approach starts from a basis of rung 
states, appropriate  for small values of the chain hopping parameter $t$,
 and we obtain 
perturbation series for various quantities in powers of $t$ up to 
typically 10 terms. The series are well behaved out to typically 
$t/t_{\perp}\sim 0.6$. but the convergence becomes problematical
beyond that point. 
Our results are complementary to both
analytical weak coupling calculations and other numerical (DMRG and QMC) results.

We have calculate the ground state energy and the triplet spin excitation energies at 
half-filling, and the excitation energies of one-hole and two-hole states relative
to the half-filled case.
 At half-filling the system is believed to be in a 
spin-gapped insulating phase\cite{bal96,noa96} for all values of the parameters 
$U,t_{\perp}$. 
Our results confirm those of Noack {\it et al.}\cite{noa96},  showing a
sharp 
 crossover at smaller $U/t_{\perp}$ from 
strongly correlated spin-liquid or Mott insulator
(as for a simple Hubbard chain) to ``band insulator'' behaviour as 
$t_{\perp}$ is increased. The spin gap becomes very small in the
spin-liquid region.
We have peformed a careful finite-size scaling analysis using 
DMRG to show that the spin gap remain finite, however, rather than scaling to zero in the
bulk limit, at least for those moderate values of $U/t_{\perp}$
where a definite statement is possible.


A single hole doped into the Hubbard ladder will propagate as a well 
defined quasiparticle, and we have compute the dispersion relation for this 
quasiparticle.
It was found that the lowest energy two-hole state is not amenable to our
series approach; but the DMRG calculations show that two holes become strongly
bound at larger $t/t_{\perp}$. This raises the question of whether
phase separation will occur in this region, as it does in the $t-J$ ladder\cite{rom00}.
This question must await future investigations.


\acknowledgments
This work forms part of a research project supported by a grant from the 
Australian Research Council. 
One of us (R.J.B.)
would like to thank Dr William Barford for co-development of one
of the DMRG codes. We would also like to thank Dr. Reinhard Noack for
sending us his DMRG results for comparison.
The computation has been performed on 
Silicon Graphics Power Challenge and Convex machines.  We thank the New 
South Wales Centre for Parallel Computing for facilities and assistance 
with the calculations. 



\begin{table}
\caption{The sixteen rung states and their energies, where $u_1=\frac{1}{2}
\sqrt{1 + U/\sqrt{U^2+16 t_{\perp}^2}}$,
$u_2=\frac{1}{2} \sqrt{1 - U/\sqrt{U^2+16 t_{\perp}^2}}$, 
$\lambda_1 = \frac{1}{2} (U-\sqrt{U^2+16 t_{\perp}^2})$,
$\lambda_2 = \frac{1}{2} (U+\sqrt{U^2+16 t_{\perp}^2})$,
and $0$ represent hole, $\uparrow$ ($\downarrow$) represent 
up (down)-spin electron, $\updownarrow$ represent an electron
pair.
}\label{tab1}
\begin{tabular}{|c|c|c|c|}
\multicolumn{1}{|l|}{State label} &\multicolumn{1}{c|}{Eigenstate}
&\multicolumn{1}{l|}{Eigenvalue} &\multicolumn{1}{c|}{Name} \\
\hline
0  &  $u_1 (\mid \uparrow \downarrow \rangle - \mid \downarrow \uparrow 
\rangle )  
- u_2 (\mid \updownarrow 0 \rangle + \mid 0 \updownarrow \rangle )$ &    
$\lambda_1$  &  singlet  \\
\hline
1  & $\frac{1}{\sqrt 2}(\mid 0 \downarrow \rangle + \mid \downarrow 0 
\rangle )$ & $-t_{\perp}$ 
 & electron-hole bonding ($S^z_{\rm tot}=-\frac{1}{2}$)  \\
\hline
2  & $\frac{1}{\sqrt 2}(\mid 0 \uparrow \rangle + \mid \uparrow 0 
\rangle )$ & $-t_{\perp}$ 
 & electron-hole bonding ($S^z_{\rm tot}=\frac{1}{2}$)  \\
\hline

3  & $\frac{1}{\sqrt 2}(\mid \updownarrow \downarrow \rangle - \mid 
\downarrow \updownarrow \rangle )$ 
  & $U-t_{\perp}$  & three-electron antibonding ($S^z_{\rm tot}=-\case 
1/2$)  \\
\hline
4  & $\frac{1}{\sqrt 2}(\mid \updownarrow \uparrow \rangle - \mid 
\uparrow \updownarrow \rangle )$ 
  & $U-t_{\perp}$  & three-electron antibonding ($S^z_{\rm tot}=\case 
1/2$)  \\
\hline

5  & $ \mid \downarrow \downarrow \rangle   $  &    $0$  &  triplet 
($S^z_{\rm tot}=-1$)  \\
\hline
6  & $\frac{1}{\sqrt 2}(\mid \uparrow \downarrow \rangle + \mid 
\downarrow \uparrow \rangle )$ & $0$  &  triplet ($S^z_{\rm tot}=0$)  \\
\hline
7  & $\mid \uparrow \uparrow \rangle  $  &    $0$  &  triplet ($S^z_{\rm 
tot}=1$)  \\
\hline

8  & $\mid 00 \rangle$  &    $0$  &  hole-pair singlet  \\
\hline

9  & $\frac{1}{\sqrt 2}(\mid \updownarrow 0 \rangle - \mid 0 
\updownarrow \rangle )$ 
  & $U$  & an electron pair and a hole singlet \\
\hline

10  & $\mid \updownarrow \updownarrow \rangle $ 
  & $2 U$  & an electron-pairs  \\
\hline

11  & $\frac{1}{\sqrt 2}(\mid 0 \downarrow \rangle - \mid \downarrow 0 
\rangle )$ & $t_{\perp}$  & electron-hole antibonding ($S^z_{\rm tot}=-
\frac{1}{2}$)  \\
\hline
12  & $\frac{1}{\sqrt 2}(\mid 0 \uparrow \rangle - \mid \uparrow 0 
\rangle )$ & $t_{\perp}$  & electron-hole antibonding ($S^z_{\rm 
tot}=\frac{1}{2}$)  \\
\hline
13  & $\frac{1}{\sqrt 2}(\mid \updownarrow \downarrow \rangle + \mid 
\downarrow \updownarrow \rangle )$ 
  & $U+t_{\perp}$  & three-electron bonding ($S^z_{\rm tot}=-\frac{1}{2}$)  
\\
\hline
14  & $\frac{1}{\sqrt 2}(\mid \updownarrow \uparrow \rangle + \mid 
\uparrow \updownarrow \rangle )$ 
  & $U+t_{\perp}$  & three-electron bonding ($S^z_{\rm tot}=\frac{1}{2}$)  
\\
\hline  
15  &  $u_2 (\mid \uparrow \downarrow \rangle - \mid \downarrow \uparrow 
\rangle )  
+ u_1 (\mid \updownarrow 0 \rangle + \mid 0 \updownarrow \rangle )$ &    
$\lambda_2$  & mixed singlet  
\end{tabular}
\end{table}

\begin{table}
%
\setdec 0.000000000000000
\squeezetable
\caption{
Nonzero coefficients of $(t/t_{\perp})^n$ for the ground state energy 
per site $E_0/N$, the spin gap $\Delta_s$, the 1-hole energy 
$\Delta_{\rm 1h}(\pi)$, and the 2-hole energy $\Delta_{\rm 2h}(\pi)$ at 
$U/t_{\perp}=8$.
}
\label{tab2}
\begin{tabular}{||c|c|c|c||l|c||}
\multicolumn{1}{||l|}{n} &\multicolumn{1}{c|}{$E_0/NJ_{\perp}$}
&\multicolumn{1}{c|}{$\Delta_s/J_{\perp}$} 
&\multicolumn{1}{c||}{$\Delta_{\rm 2h}/J_{\perp}$}
 & \multicolumn{1}{l|}{n} &\multicolumn{1}{c||}{$\Delta_{\rm 
1h}/J_{\perp}$} \\
\hline
  0 &\dec -2.360679774998$\times 10^{-1}$ &\dec  4.721359549996$\times 
10^{-1}$ &\dec  4.721359549996$\times 10^{-1}$ & 0 &\dec 
-5.278640450004$\times 10^{-1}$ \\
  2 &\dec -8.944271909999$\times 10^{-2}$ &\dec -7.478019326001$\times 
10^{-1}$ &\dec  3.577708764000$\times 10^{-1}$ & 1 &\dec -1.447213595500      
\\
  4 &\dec -7.055810863547$\times 10^{-2}$ &\dec  2.984040157758$\times 
10^{-1}$ &\dec  2.117784192484$\times 10^{-1}$ & 2 &\dec 
-5.681892697507$\times 10^{-2}$ \\
  6 &\dec -5.448947627757$\times 10^{-2}$ &\dec  2.126553951831$\times 
10^{-1}$ &\dec -1.022223221785                & 3 &\dec  
4.190688837075$\times 10^{-1}$ \\
  8 &\dec -3.092587547718$\times 10^{-2}$ &\dec  1.030100155673$\times 
10^{-1}$ &\dec -1.054110318250$\times 10^{1}$ & 4 &\dec 
-2.186544916675$\times 10^{-2}$ \\
 10 &\dec  5.006769301333$\times 10^{-3}$ &\dec -2.047688788511$\times 
10^{-2}$ &\dec -8.623391111290$\times 10^{1}$ & 5 &\dec  
7.778879476658$\times 10^{-1}$ \\
 12 &\dec  4.598641784004$\times 10^{-2}$ &\dec -2.168399847938$\times 
10^{-1}$ &\dec -6.784316243528$\times 10^{2}$ & 6 &\dec -1.450073605871      
\\
 14 &\dec  7.235038860679$\times 10^{-2}$ &                                    
&                                   & 7 &\dec  4.979933826800      \\
 16 &                           &                           &                           
& 8 &\dec -1.393584028913$\time 10^{1}$ \\
 18 &                           &                           &                           
& 9 &\dec  4.115238962605$\time 10^{1}$ \\
\end{tabular}
\end{table}

\begin{table}
\caption{
Comparison between exact and DMRG results for the spin gap 
$\Delta_s / t_\perp$ for various lattice sizes $N_{\rm rung}$ for the 
case of $U = 0$, $t / t_\perp = 0.49$. The DMRG results are obtained 
using the odd-rung algorithm, retaining around $m = 700$ states per 
block, and including the ground and first excited states in the density 
matrix with equal weights. The $N_{\rm rung} = \infty$ DMRG result is 
obtained by extrapolating the finite $N_{\rm rung}$ DMRG results over the
range $63 \leq N_{\rm rung} \leq 81$, assuming that the corrections to 
the bulk result scale as $1 / N_{\rm rung}^2$.
}
\begin{tabular}{l|ll}
$N_{\rm rung}$
    &    DMRG    &    Exact   \\
\hline
3   &  0.614071  &  0.614071  \\
9   &  0.136013  &  0.135929  \\
19  &  0.064140  &  0.064131  \\
37  &  0.046701  &  0.046694  \\
55  &  0.043092  &  0.043083  \\
67  &  0.042103  &  0.042091  \\
81  &  0.041456  &  0.041438  \\
$\infty$
    &  0.040055  &  0.040000
\end{tabular}
\label{table_3}
\end{table}

\begin{table}
\caption{
DMRG convergence of the spin gap $\Delta_s / t_\perp$ with $m$, the 
number of states retained per block, for the odd-rung algorithm, where 
the ground and first excited states are included in the density matrix 
with equal weights. The parameters used are $U / t_\perp = 1$, 
$t / t_\perp = 0.6$.
}
\begin{tabular}{c|cccccc}
$N_{\rm rung}$ & $m = 200$ & $m = 370$ & $m = 580$ & $m = 750$ & $m = 1200$
& $m = 1500$ \\
\hline
3  & 0.2371380 & 0.2371380 & 0.2371380 & 0.2371380 & 0.2371380 & 0.2371380 \\
7  & 0.1115510 & 0.1115030 & 0.1115060 & 0.1115040 & 0.1115010 & 0.1115010
\\
13 & 0.0675853 & 0.0673585 & 0.0672920 & 0.0672718 & 0.0672626 & 0.0672597
\\
19 & 0.0475570 & 0.0475439 & 0.0475508 & 0.0475459 & 0.0475420 & 0.0475383
\\
25 & 0.0382940 & 0.0383589 & 0.0383977 & 0.0384008 & 0.0384094 & 0.0384077
\\
31 & 0.0333642 & 0.0335102 & 0.0336259 & 0.0336437 & 0.0336687 & ---
\\
35 & 0.0314530 & 0.0317155 & 0.0319372 & 0.0319716 & 0.0320189 & ---
\end{tabular}
\label{table_4}
\end{table}



\begin{figure}[p] 
\caption{
The ground state energy per site as function of $t/t_{\perp}$ for 
$U/t_{\perp}=0,0.5,1,2,4,8$. The solid lines are extrapolations of
the series using different integrated differential approximants, 
while the points connected by dashed lines are the results of DMRG 
calculations. Also shown are the exact results for $U=0$, and the results
of a variational approach\protect\cite{kim99} for $U/t_{\perp}=4,8$ and 
$t/t_{\perp}=0.5,1$ 
(full circle points).
}
\label{fig_2}
\end{figure}

\begin{figure}[p] 
\caption{
Spin excitation spectra  for $t/t_{\perp}=0.5$ and various 
$U/t_{\perp}$, obtained from series expansions.
}
\label{fig_3}
\end{figure}

\begin{figure}[p] 
\caption{
The spin gap $\Delta_s$ versus $t / t_{\perp}$ for various 
$U / t_{\perp}$. The dashed lines  are extrapolations of the 
series using different integrated differential approximants, while 
the points connected by solid lines are the results of DMRG 
calculations. The filled circles (crosses) denote results obtained from 
the even-rung (odd-rung) algorithm.
}
\label{fig_4}
\end{figure}

\begin{figure}[p] 
\caption{
Finite-size scaling of the spin gap $\Delta_s$ in the noninteracting 
case ($U = 0$) for $t / t_\perp = 0.49$ (a),
$t / t_\perp = 0.51$ (b) and $t / t_\perp = 0.55$ (c). These
are exact results for open lattices.
}
\label{fig_5a}
\end{figure}

\begin{figure}[p] 
\caption{
Finite-size scaling of the spin gap $\Delta_s$ for 
$U / t_{\perp} = 1$ 
and $t / t_{\perp} = 0.54, 0.6$, obtained from DMRG calculations. For 
$t/t_{\perp}=0.6$, the results from the two different DMRG algorithms: 
odd-rungs (crosses) and even-rungs (full dots), retaining 750 and 800 
states per block respectively, are compared. The inset shows the FSS in 
the $U / t_\perp = 1$, $t / t_{\perp} = 0.8$ case using the odd-rung
DMRG algorithm with $m = 1200$.
}
\label{fig_5}
\end{figure}

\begin{figure}[p] 
\caption{
Phases and critical lines for the Hubbard ladder at half-filling, in the 
plane of $U/t_{\perp}$ versus $t/t_{\perp}$ (see text). The phase 
boundary between spin liquid and band insulator is obtained 
approximately by determining the position where the spin gap
vanishes according to the series. The region that two holes bind (do not bind) is  marked by 
B(NB). The region in which DMRG fails to determine whether the system
has a spin  gap, due the irregular finite size scaling, is marked by F.
Also marked is the region that the 1-hole gap is positive/negative.
}
\label{fig_6}
\end{figure}

\begin{figure}[p] 
\caption{
Excitation spectra for one-hole bonding states for $t/t_{\perp}=0.5$ and 
various $U/t_{\perp}$, obtained from series expansions. Also shown are 
the results of the local rung approximation\protect\cite{end96} at 
$U/t_{\perp}=4$ (dashed line).
}
\label{fig_7}
\end{figure}

\begin{figure}[p] 
\caption{1-hole ``quasiparticle gap" as a function of $t/t_{\perp}$ for 
various $U/t_{\perp}$. The dashed lines are the extrapolations of
the series using the different integrated differential approximants, 
while the points connected by solid lines are the results of DMRG 
calculations. Also shown are the exact results for the case $U=0$.
}
\label{fig_8}
\end{figure}

\begin{figure}[p] 
\caption{1-hole ``quasiparticle gap'' $\Delta_{1h}(\pi)$  for 
$U/t_{\perp}=1$ and $t/t_{\perp}=0.5,0.52,0.54,0.6$, as a function of 
$1/N^2_{\rm rung}$, obtained from DMRG calculations.
}
\label{fig_9}
\end{figure}

%
%

\begin{figure}[p] 
\caption{Binding energy $E_b$ of two holes versus $t/t_{\perp}$ for various 
$U/t_{\perp}$, obtained from DMRG calculations.
}
\label{fig_12}
\end{figure}

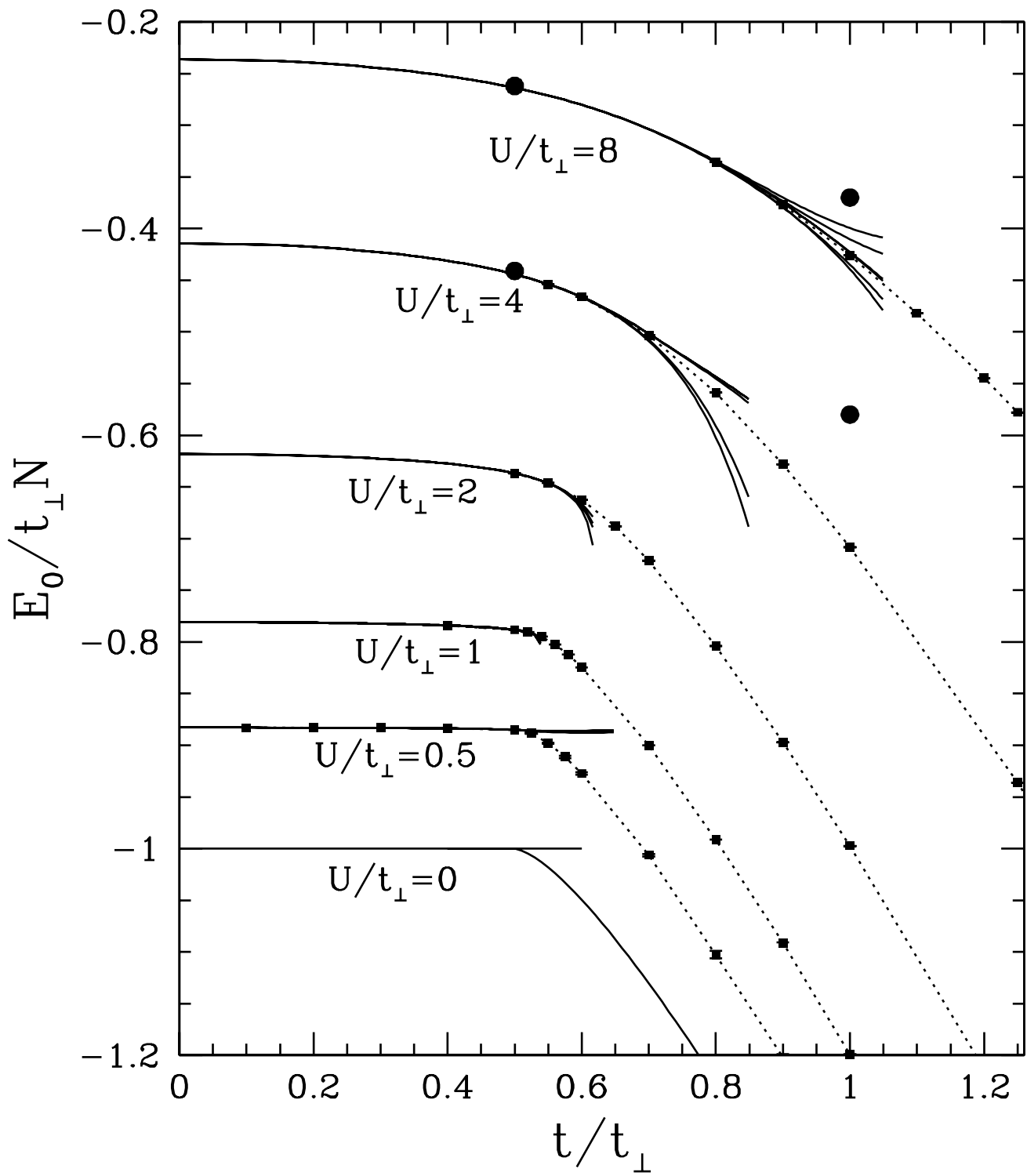
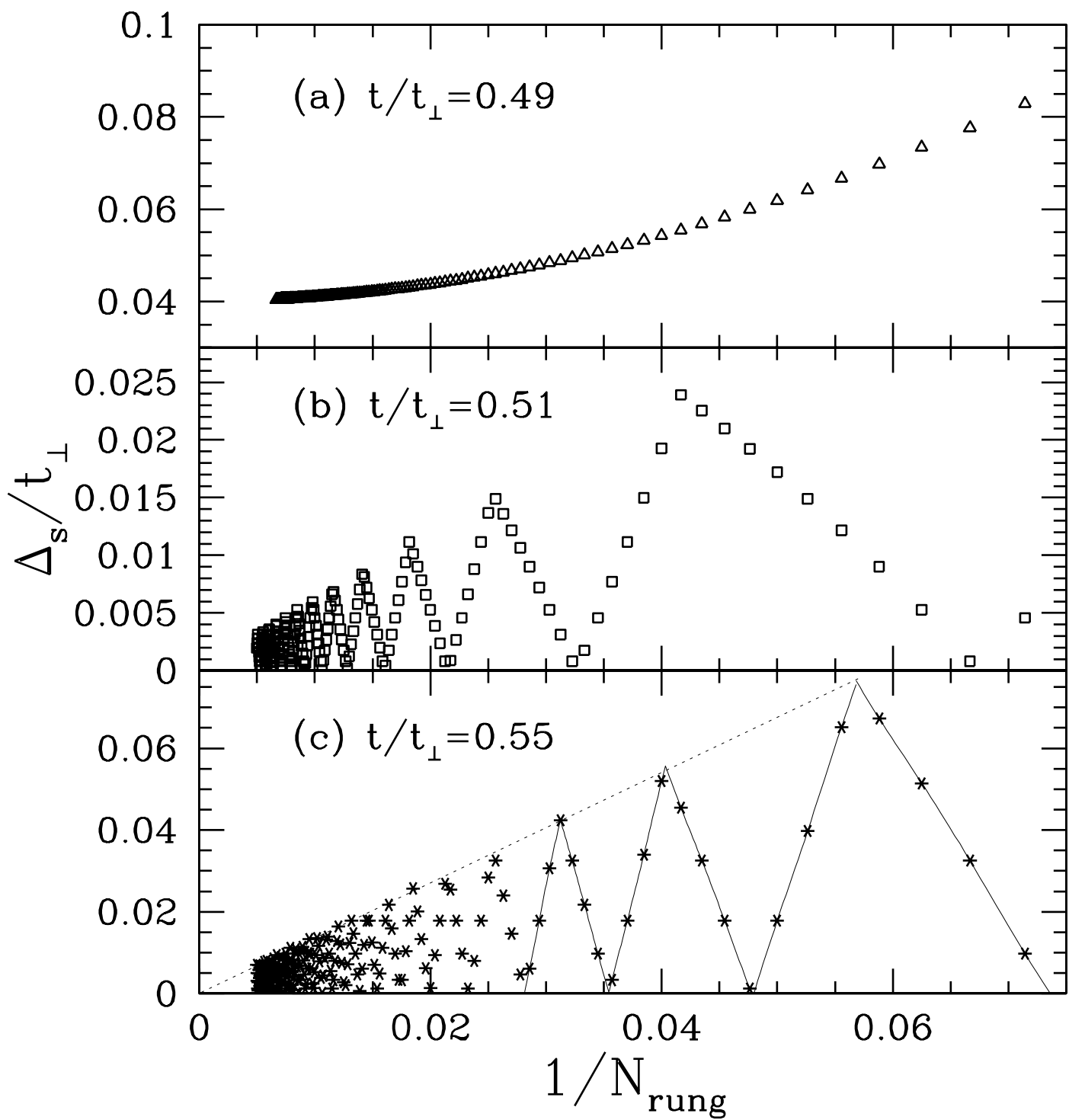
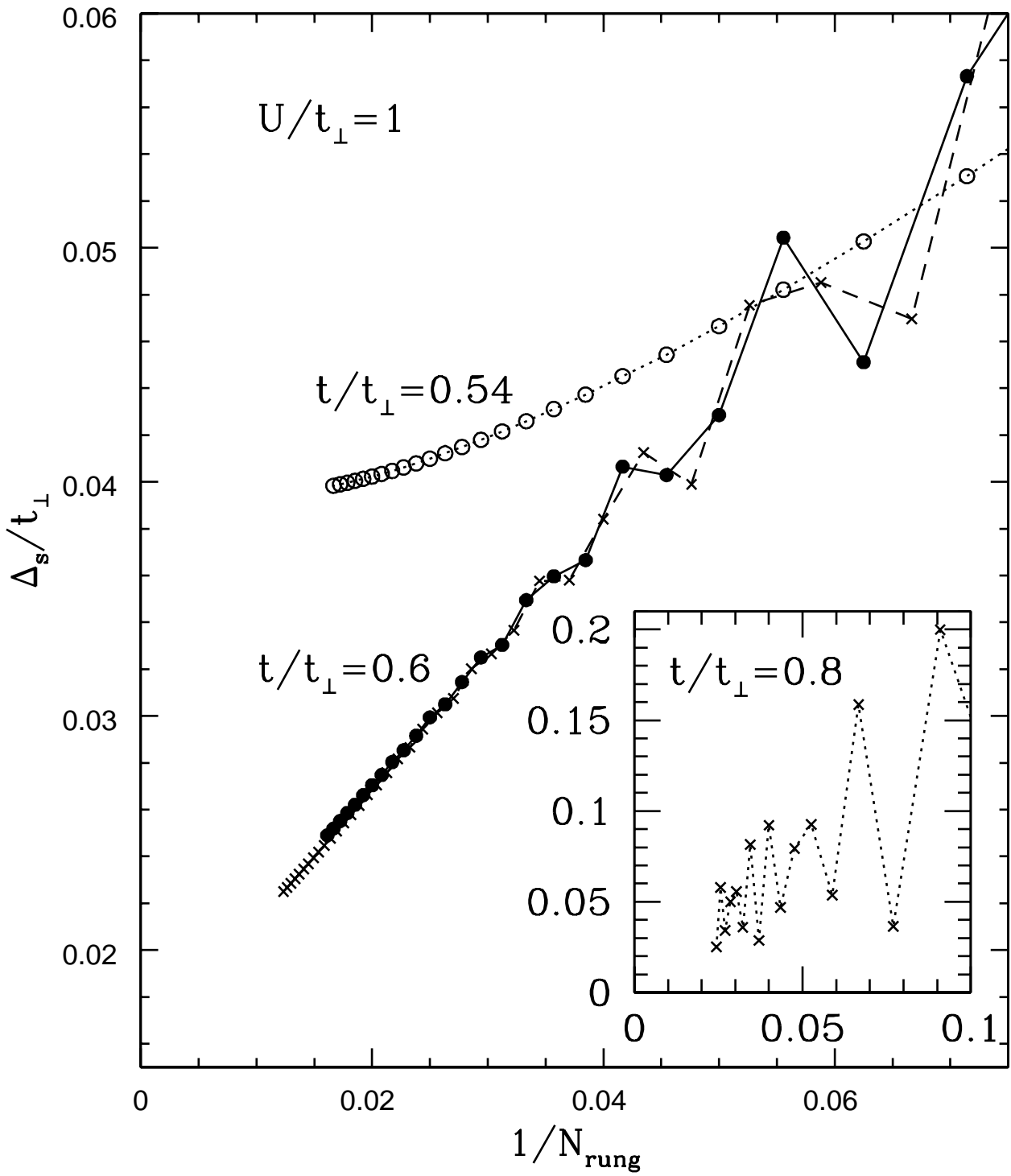
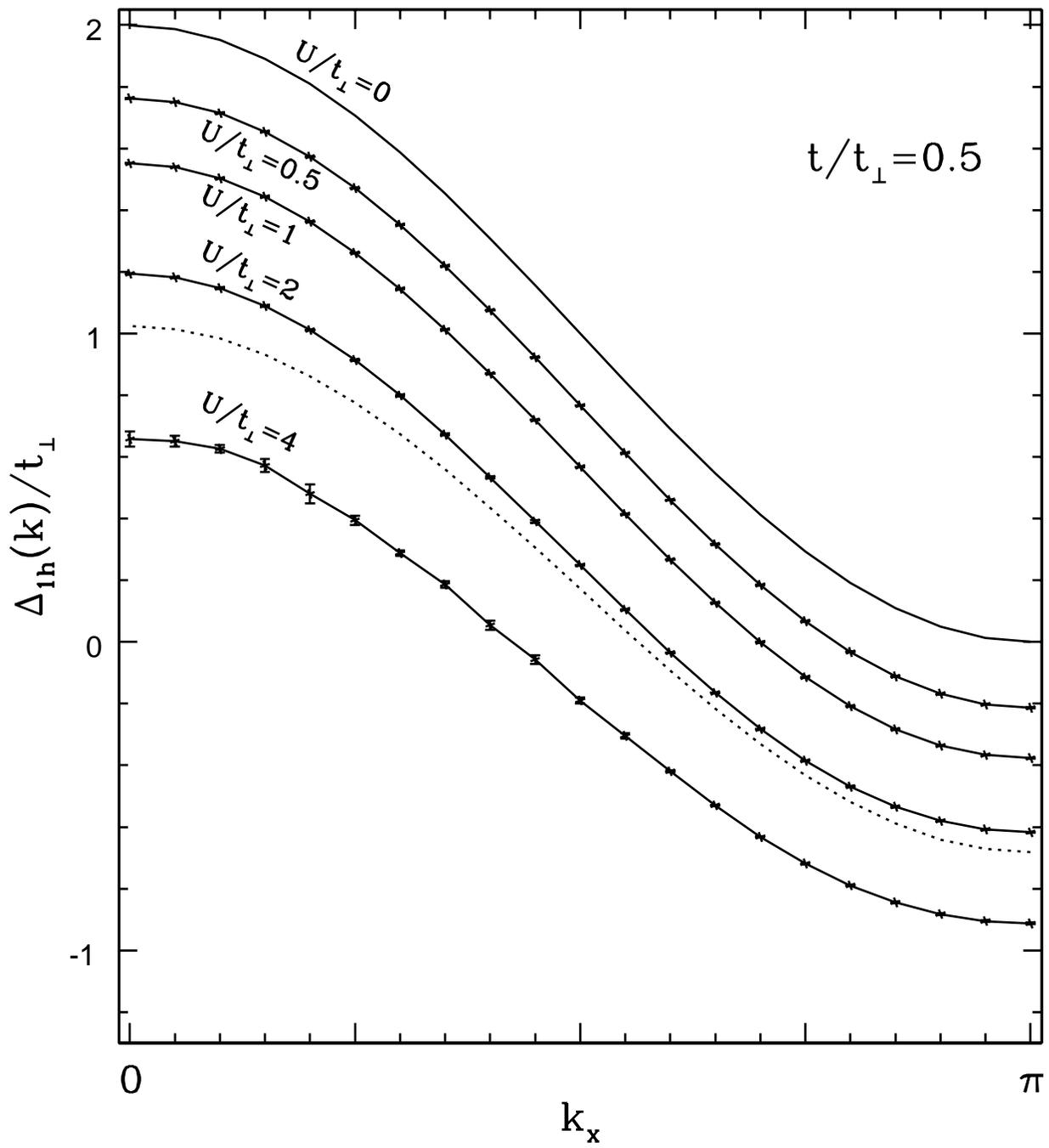
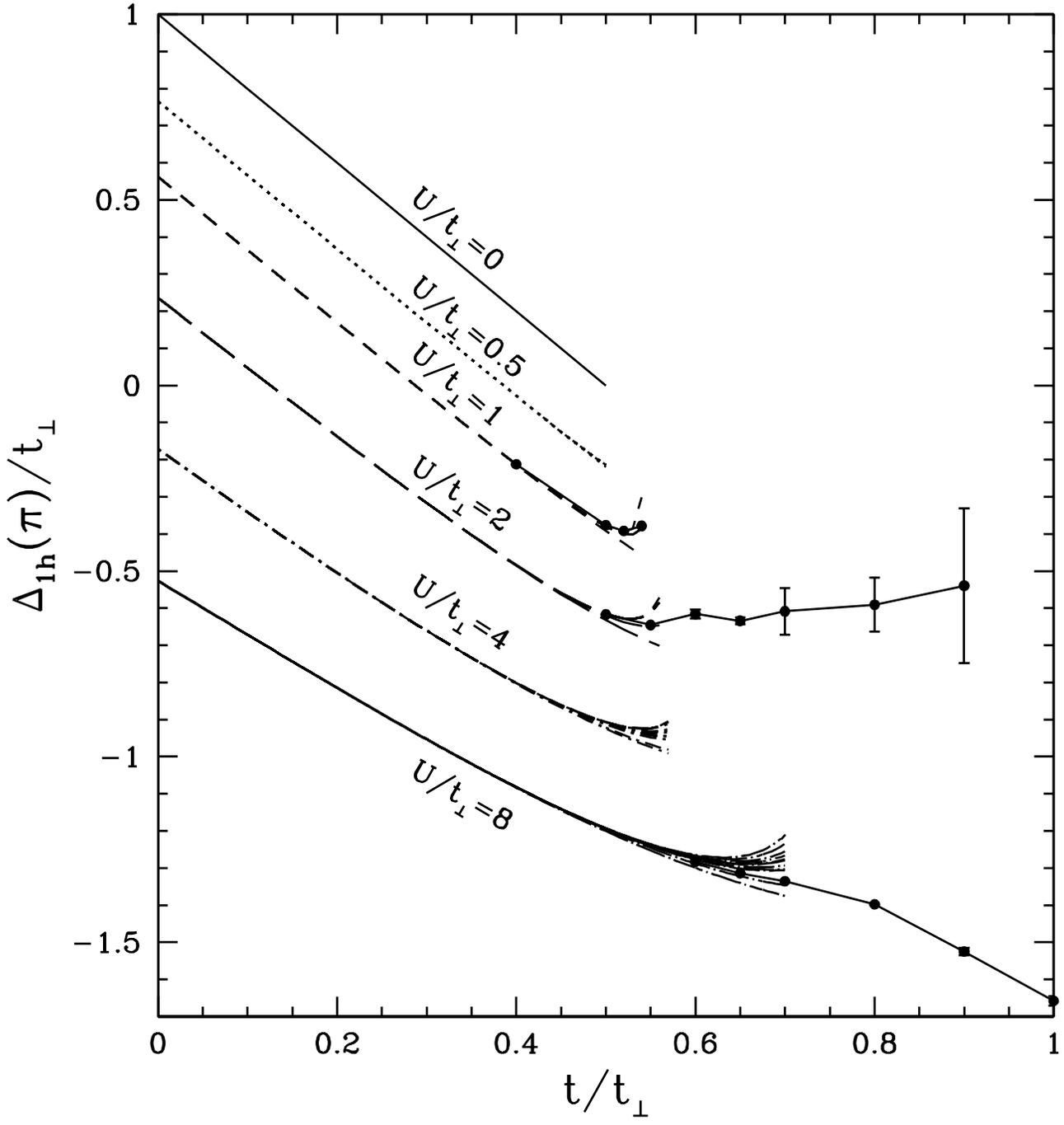
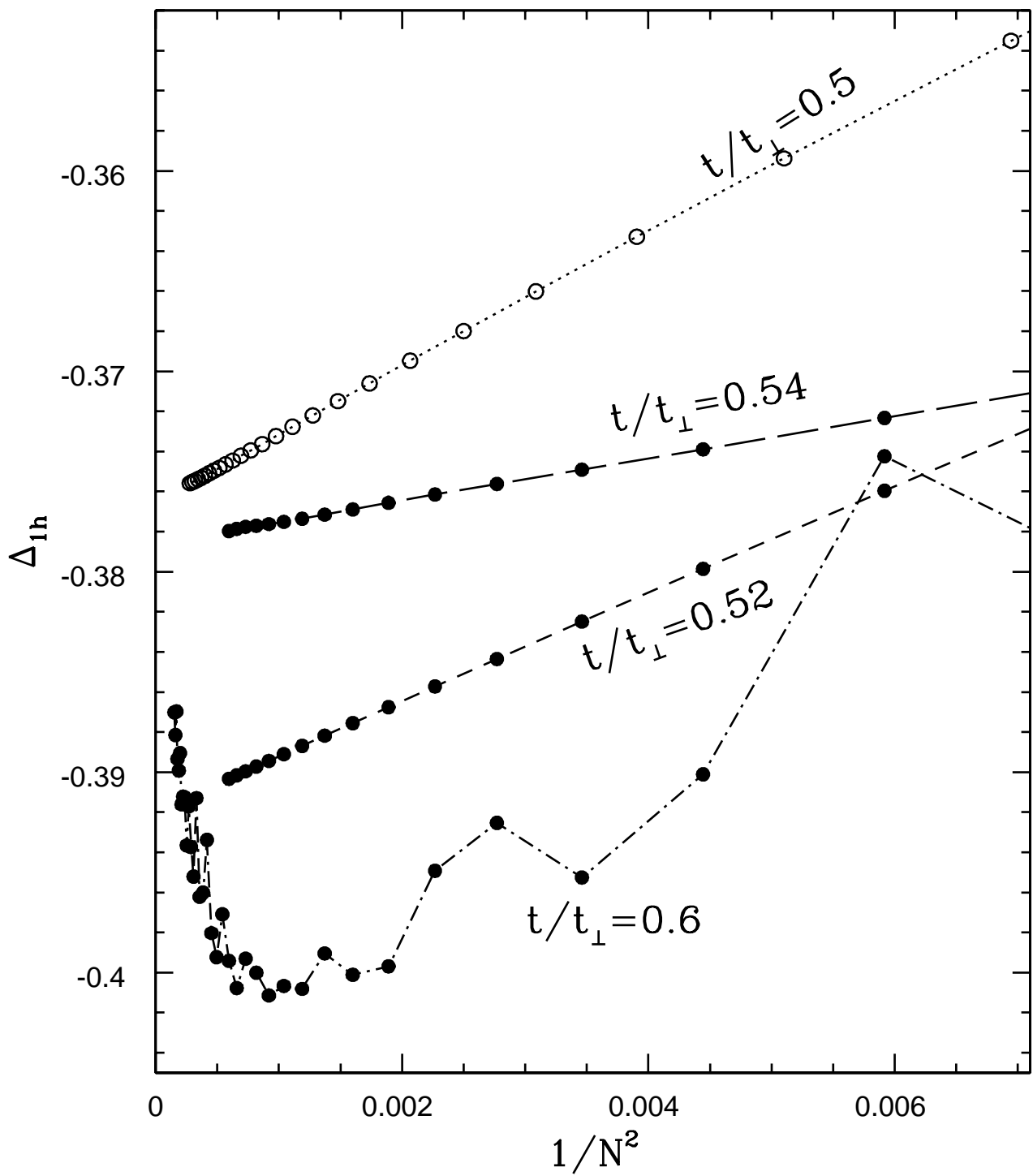
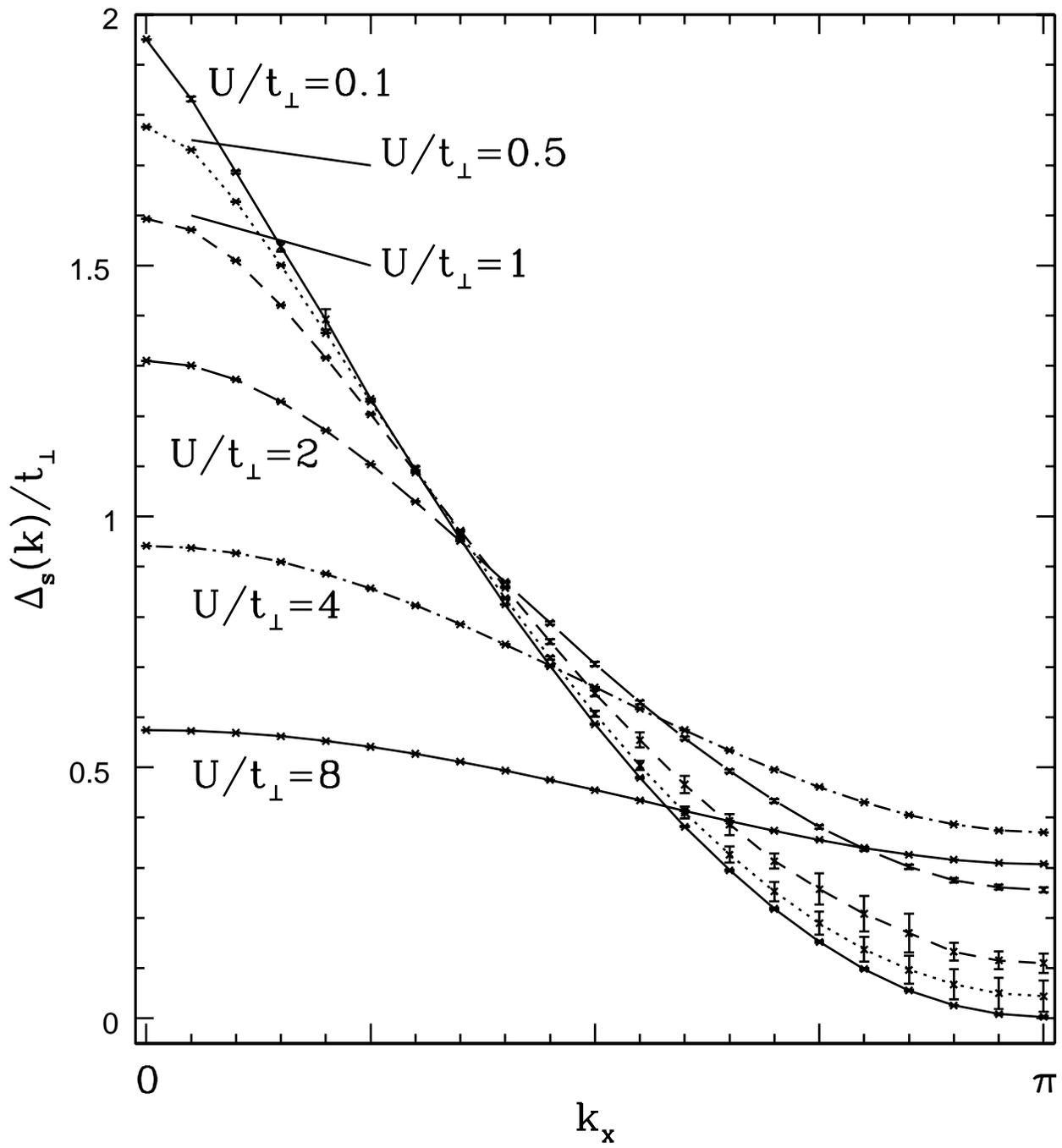
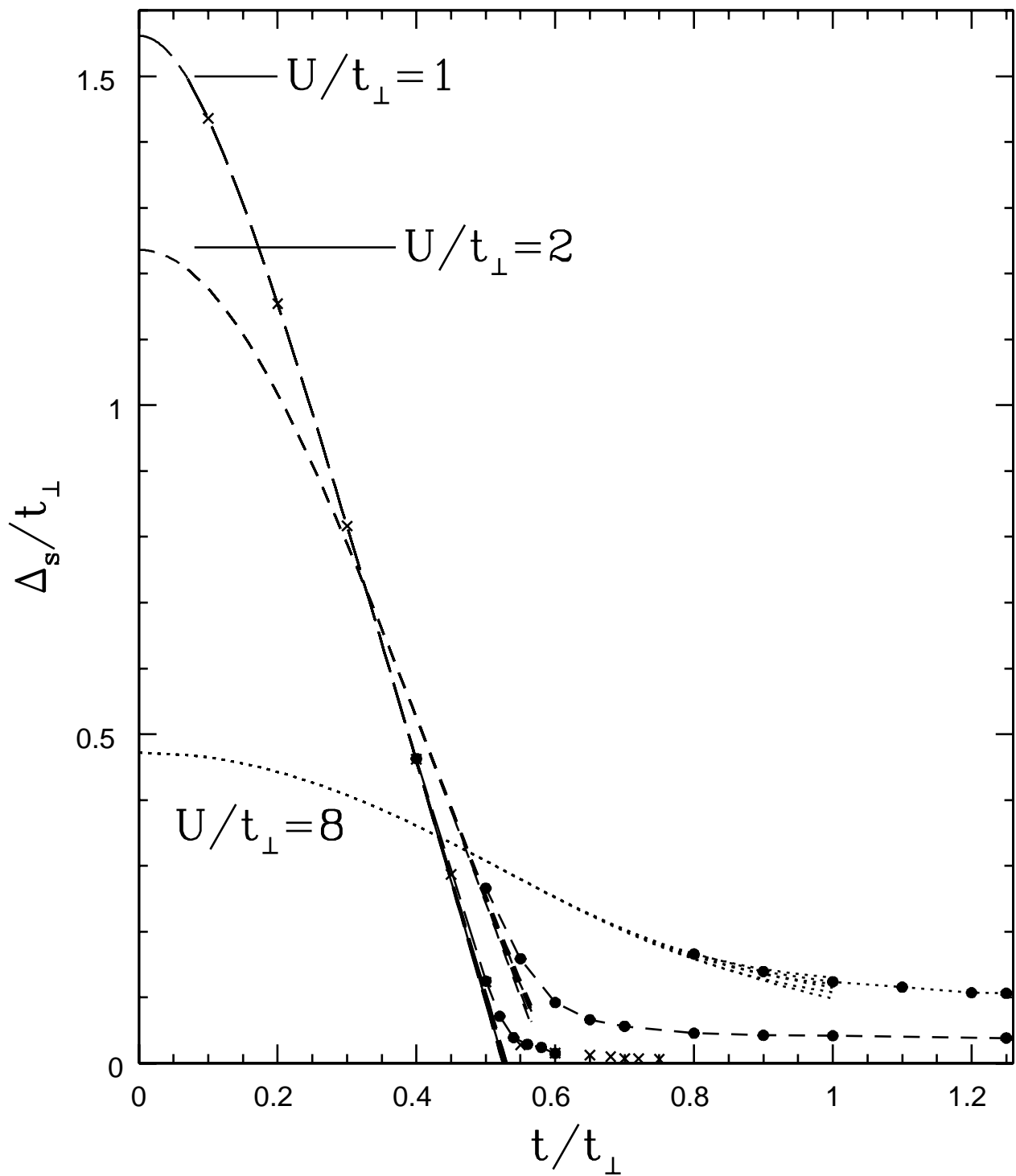
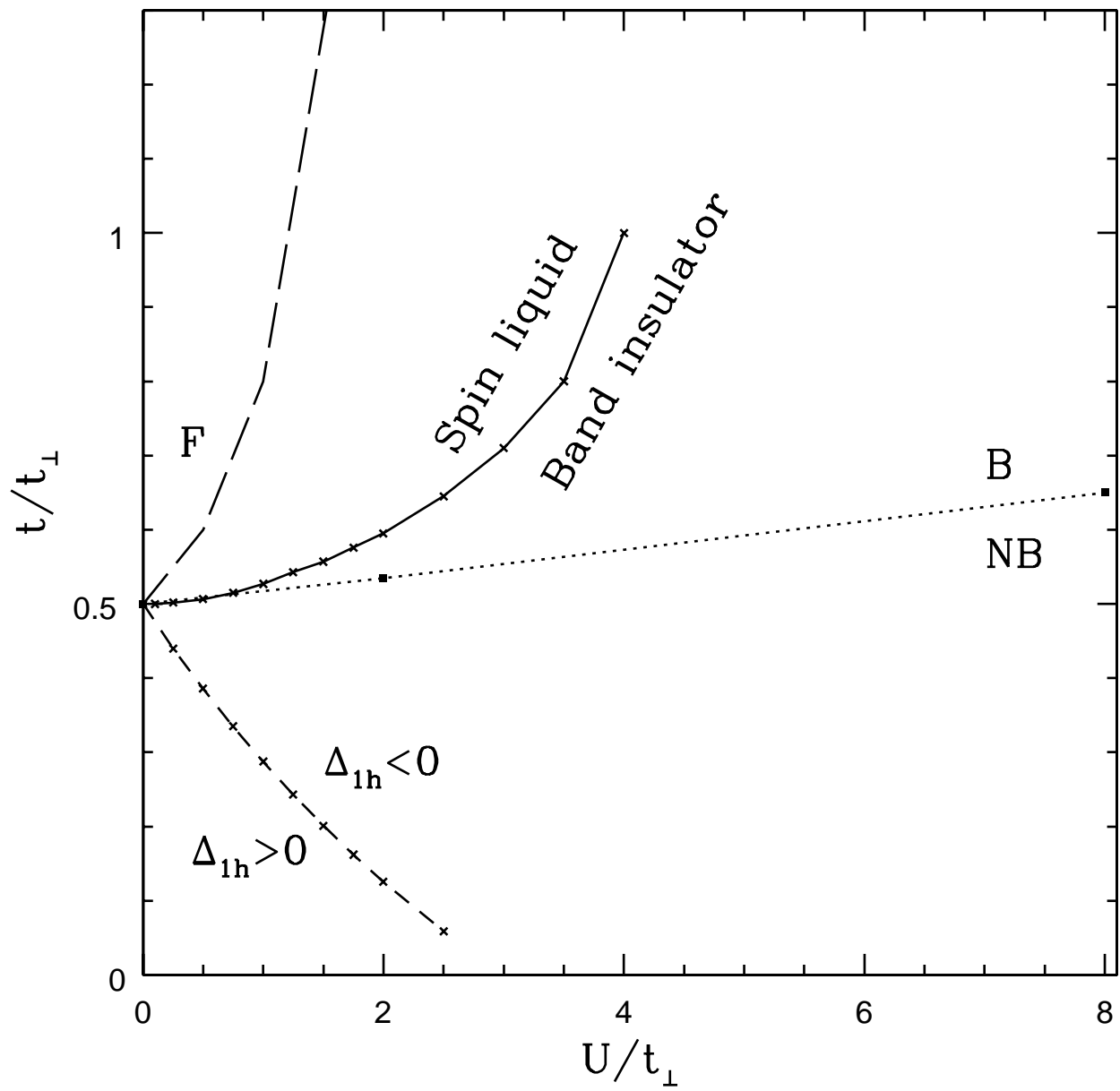
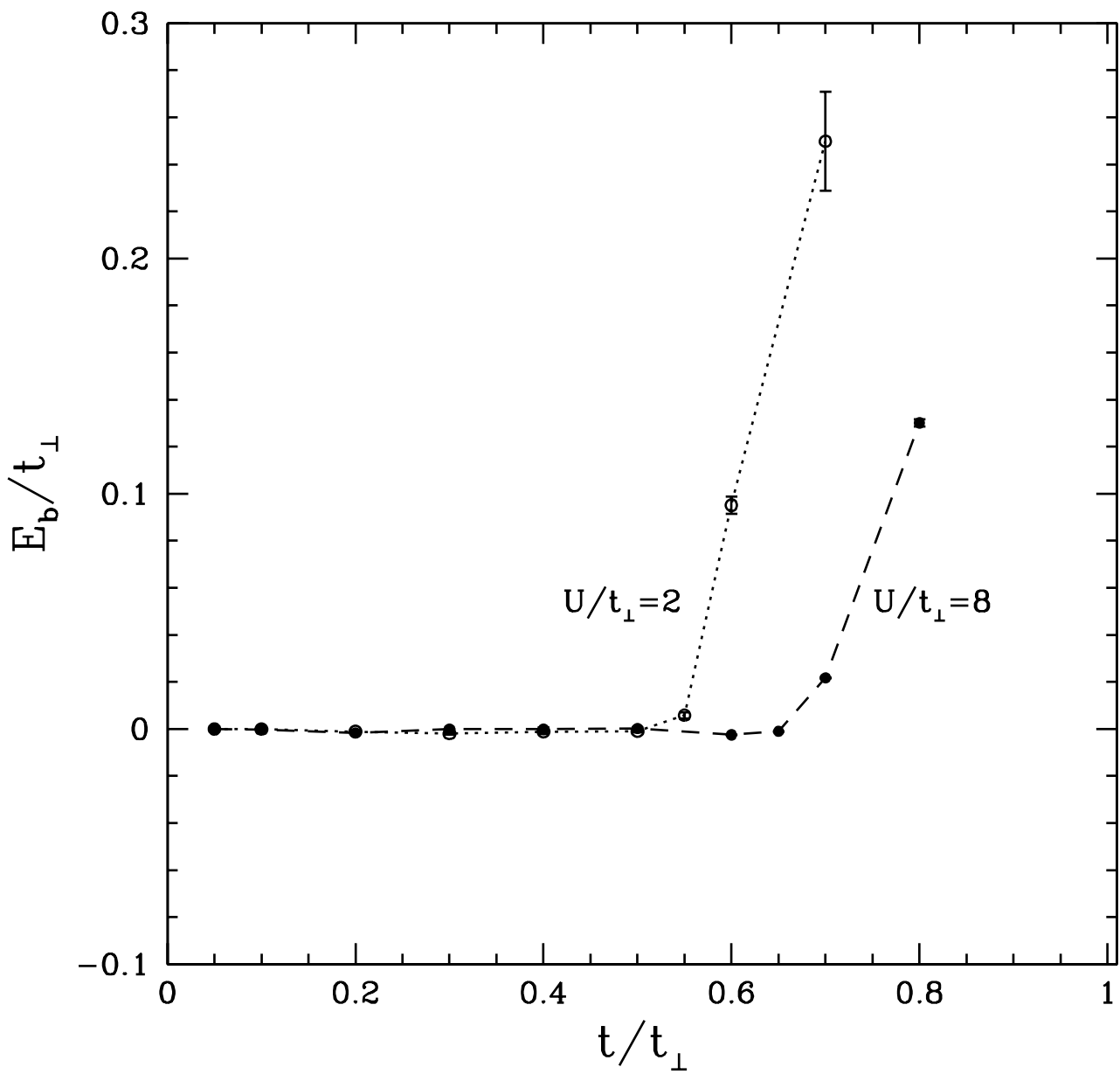
\end{document}